\documentclass[a4paper,12pt]{article}  
  
\usepackage[a4paper,tmargin=3truecm,bmargin=3truecm,rmargin=2.5truecm,  
lmargin=2.5truecm,twoside,verbose=true]{geometry}  
\usepackage{amsmath}  
\usepackage{amssymb,epsfig}  
  
\usepackage{stmaryrd,wasysym,upgreek,latexsym}  
\usepackage{lscape} 
\usepackage[vflt]{floatflt}  
\usepackage{afterpage}

\def\NN{{\mathbb N}}  
\def\CC{{\mathbb C}}  
  
\def\NN{{\mathbb N}}

\newcommand{\beq}{\begin{equation}}  
\newcommand{\eeq}{\end{equation}}  
\newcommand{\beqa}{\begin{eqnarray}}  
\newcommand{\eeqa}{\end{eqnarray}}  
\newcommand{\noi}{\noindent}

\newcommand{\tx}{\widetilde{x}}  
  
\newcommand{\resetequ}{\setcounter{equation}{0}}

\begin{document}  
  
\title{Vacuum configurations for renormalizable
  non-commutative scalar models}  
\author{  
A. de Goursac${}^{(1), (2)}$, A. Tanasa${}^{(1), (3)}$ and  
J-C. Wallet${}^{(1)}$\footnote{e-mail:  
  axelmg@melix.net, adrian.tanasa@ens-lyon.org, jean-christophe.wallet@th.u-psud.fr}\\  
${}^{(1)}$Laboratoire de Physique Th\'eorique, B\^at.\ 210, CNRS UMR 8627\\  
    Universit\'e Paris XI,  F-91405 Orsay Cedex, France\\  
${}^{(2)}$Mathematisches Institut der Westf\"alischen  
  Wilhelms-Universit\"at \\Einsteinstra\ss{}e 62, D-48149 M\"unster,  
  Germany \\  
${}^{(3)}$ Dep. Fizica Teoretica, Institutul de Fizica si Inginerie Nucleara  
    H. Hulubei,\\  
    P. O. Box MG-6, 077125 Bucuresti-Magurele, Romania  
}  
\maketitle  
  
\begin{abstract}  
In this paper we find non-trivial vacuum states for the renormalizable  
non-commutative $\phi^4$ model. An associated linear sigma  
model is then considered. We further investigate the corresponding spontaneous symmetry breaking.     
\end{abstract}  
  
  
\pagebreak  
  
\section{Introduction}  
\resetequ  
  
Non-commutative quantum field theory (for a general review see \cite{DN} or  
\cite{Szabo}) - that is field theory based on non-commutative geometry (see  
\cite{book-connes} for a general review) - is nowadays one of the most appealing candidates for New Physics  
beyond the Standard Model. Moreover, non-commutative field theory can be seen  
as an effective regime of string theory (see \cite{string1,  
  string2}). From a different point of view, non-commutativity is well adapted  
for the description of physics in the presence of a background field, like for  
example the fractional quantum Hall effect \cite{hall1, hall2, hall3}.   
  
Non-commutative physics is known to suffer from a new type of divergence, the  
UV-IR mixing, divergence which is responsible for the non-renormalizability of  
the models. This difficulty was overtaken for scalar $\Phi^4$ models by the  
introduction of a new harmonic term in the action - the Grosse-Wulkenhaar  
model.  
The model was proven to be  
renormalizable at any order in perturbation theory \cite{GW1, GW2, GW5,   
  GW3, GW4}. Moreover, the parametric representation was introduced \cite{param1}  
and then the dimensional regularization and renormalization were performed   
\cite{dimreg}. Let us also emphasize here that the Hopf algebra  
description of this type of renormalization was given in  
\cite{hopf}. Moreover, let us also stress here on the fact that it was recently shown  
\cite{ultimul} that this type of action can be interpreted from the spectral  
action (for latest developments see \cite{spectral}) point of view.

The Grosse-Wulkenhaar model was however proven to have a better flow behavior  
with respect to the commutative $\phi^4$ model. Indeed, in \cite{beta1,  
  beta23} and \cite{beta} was proven that this model does not present a Landau  
ghost; let us recall that this was not the case for the commutative model.   
  
Another improvement with respect to commutative scalar quantum field theory is  
that a constructive version (for a general review see \cite{carte}) is within  
reach \cite{constructiva1, constructiva2}.

\medskip  
  
In this paper we first obtain vacuum states which, because of the presence of this new  
harmonic term, must be non-trivial functions of the space-time  
position $x$.   
Note that a somewhat similar conclusion regarding non constant vacuum was also
obtained recently in  \cite{ultimul}. We then analyze the issue of spontaneous symmetry breaking  
 for a non-commutative analog of the linear sigma model with harmonic term at the classical level.  
 The model we consider here is  a non-commutative linear sigma model  
based on a set of $N$ scalar fields but in the presence of  harmonic  
 terms for each of these scalars. 

Note that the Goldstone  
theorem for the non-commutative linear sigma model 
without harmonic term 
was already investigated up to one-loop \cite{gold1, gold11} and  
two-loops \cite{gold2}. Other investigations regarding different  
non-commutative models were  
done in \cite{higgs, stripe, altii1, altii2}. 
Within these models it was found that the situation for the Goldstone theorem,
followed rather closely the features of the commutative case. In the present
case we find that the situation is much more involved.


  
\medskip  
  
The paper is organized as follows. In section $2$ we give some notations  
and conventions and we introduce the Grosse-Wulkenhaar model as well as some  
existing results. In the third section we find vacuum states $v(x)$ analyzing  
under what conditions they are   
solutions of the equations of motion.  Then we investigate the  
issue of spontaneously symmetry breaking for the  
linear sigma model. Finally, in the             section $4$ our concluding remarks and  
discussions are given.

\section{Notations and conventions. The Grosse-Wulkenhaar model}  
\resetequ  
  
We first collect the basic ingredients on the Moyal algebras (see for example \cite{matrix1,
  moyal2, raimar} or \cite{gauge4} and references therein). 
We consider a 
$D-$dimensional Moyal algebra $\mathcal{M}$ which can be conveniently defined
using the following relation 
\beqa  
\label{2D}  
[x^\mu, x^\nu]_\star=i \Theta^{\mu \nu},  
\eeqa  
\noi  
where $[a, b]_\star = a\star b - b\star a$
 and the skewsymmetric 
matrix $\Theta$ is given by  
\begin{eqnarray}  
\label{theta}  
  \Theta=   
  \begin{pmatrix}  
    \begin{matrix} 0 &-\theta \\   
      \hspace{-.5em} \theta & 0  
    \end{matrix}    &&     0  
    \\   
    &\ddots&\\  
    0&&  
    \begin{matrix}0&-\theta\\  
      \hspace{-.5em}\theta & 0  
    \end{matrix}  
  \end{pmatrix}.  
\end{eqnarray}  
\noi  
The Moyal product of two functions   
$f$ and $g$   
can be defined by  
\beqa  
\label{moyal-product}   
 (f\star g)(x)=  
\frac{1}{\pi^{D}|\det\Theta|}\int d^{D}yd^{D}z\,f(x+y)  
  g(x+z)e^{-2\imath y\Theta^{-1}z}\; .  
\eeqa  
We will mainly consider the cases $D=2$ and $D=4$.  
Let us now list some useful formulas which will be used in the calculations.
The tracial and cyclicity relations are given by
\beqa  
\label{simplu}  
\int d^D x\ f(x)\star g(x)&=&\int d^D x\ f(x) g(x),\nonumber\\  
\int d^D x\ f(x)\star g(x)\star h(x)&=&\int d^D x\ h(x)\star f(x) \star g(x)  
\eeqa  
for any functions $f, g$ and $h$. Furthermore let
\beqa  
\label{xtilde}  
\widetilde x = 2 \Theta^{-1} x.  
\eeqa  
Note that this vector is known to play a crucial role in the construction of
canonical gauge invariant connections \cite{ gauge4, gauge1, gauge2, 
  gauge5}. 
One has
\beqa  
\label{ajutor}  
\partial_\mu \phi &=& - \frac i2 [\widetilde x_\mu, \phi]_\star,\nonumber\\  
\widetilde x_\mu \phi &=& \frac 12 \{\widetilde x_\mu, \phi \}_\star,  
\eeqa   
where $ \{a, b \}_\star=a\star b + b\star a$. Note that \eqref{theta} and \eqref{xtilde} lead to  
\beqa  
\label{int}  
\partial^\mu \widetilde x_\mu =0  
\eeqa  
and that \eqref{ajutor} can be rewritten as  
\beqa  
\label{ajutor2}  
\tx_\mu \star f &= &\tx_\mu f +i \partial_\mu f,\nonumber \\  
f\star \tx_\mu&=& \tx_\mu f -i \partial_\mu f.  
\eeqa  
  
The Moyal space, as linear space of infinite dimension, admits a particular   
base, the matrix base (for more details see for example \cite{matrix1,
  matrix2} and references within). This base involves an infinite set of
Schwartz functions
   which for $D=2$ can be indexed by two  
natural numbers $m$ and $n$, namely $f_{mn}(x)$. Some relevant properties are
given in the Appendix.

\medskip  
  
  
We consider the Euclidean action for the Grosse-Wulkenhaar model \cite{GW1,
  GW2} and its complex-valued version, respectively given by
\beqa  
\label{act-real}  
S[\phi]=\int d^Dx\big(\frac 12\partial_\mu\phi\star\partial_\mu\phi  
+\frac{\Omega^2}{2} (\widetilde{x}_\mu\phi)\star(\widetilde{x}_\mu\phi)  
-\frac{\mu^2}{2}\phi\star\phi+\lambda\,  
\phi\star\phi\star\phi\star\phi\big).\ \  
\eeqa  
and
\beqa  
\label{act-complex}  
S_C[\phi]=\int d^Dx\big(\partial_\mu\phi^\dag\star\partial_\mu\phi  
+\Omega^2 (\widetilde{x}_\mu\phi)^\dag\star(\widetilde{x}_\mu\phi)  
-\mu^2\phi^\dag\star\phi+\lambda\,  
\phi^\dag\star\phi\star\phi^\dag\star\phi\big).\ \  
\eeqa  
where one considers a negative mass parameter, $-\mu^2$.   

  
Let us emphasize here that the interest on the analyze of a real field $\phi$  
comes basically from the possible insights with respect to the study of  non-commutative  
gauge theories. 
Recently, potential candidates for renormalizable gauge theories on Moyal
spaces have been singled out in \cite{gauge1} and \cite{gauge2} (see also
\cite{gauge4} and \cite{gauge5}). Although
such a construction is a necessary step towards the elaboration of
renormalizable gauge theory, it has been recognized that these candidates have
a non-trivial vacuum \cite{ultimul, gauge1, gauge2}. Its explicit
determination is the next challenging problem that must be overcome in order
to build a meaningful perturbative expansion that could be used to check
renormalizability.


Recall that these actions are covariant under the  Langmann-Szabo duality
\cite{ls}  duality  
which relates the IR and UV regions. Throughout this paper we mainly assume
$\Omega=1$ (the point where the actions \eqref{act-real} and
\eqref{act-complex} become invariant under this Langmann-Szabo
duality). Moreover, the value $\Omega=1$ is stable under the flows of the
renormalization group \cite{beta1, beta23, beta}.


  
  
  
 

In the following we will determine and study a class of non-trivial minima for
the real-valued field theory \eqref{act-real}, using the matrix space
formalism. It turns out that non-trivial ({\it i. e.} non-constant) vacua
occur generically for  \eqref{act-real} and \eqref{act-complex} whenever
$\Omega\ne 0$. 

\section{Vacuum configurations; spontaneously symmetry breaking}  
\resetequ  
  
 
\subsection{The complex-valued scalar field theory}  
  
Let us first treat the case of the complex field. The equation of motion
obtained from \eqref{act-complex} is  
\beqa  
\label{eqmvt-complex}  
-\partial^2\phi+\Omega^2\widetilde{x}^2\phi-\mu^2\phi+2\lambda\phi\star\phi^\dag\star\phi=0.  
\eeqa  
 
From \eqref{eqmvt-complex} it can be easily seen that the harmonic term prevents
a constant non-zero field to satisfy the equation of motion. This implies that 
constant vacua are forbidden, contrary to what happens in commutative models
as well as in non-commutative models with $\Omega=0$. Note that, as already
stated in the introduction, a similar observation was done in \cite{ultimul}. 
Moreover, note  that a specific type of non-constant configuration $v(x)$, stripe phases,  
were analyzed in a different context in a non-commutative framework in  
\cite{stripe}.  
  
Using \eqref{ajutor}, equation \eqref{eqmvt-complex} can be rewritten as   
\beqa  
\label{eqmvt-complex2}  
\frac 14 (1+\Omega^2) (\tx^2 \star \phi + \phi\star \tx^2) - \frac 12  
(1-\Omega^2) \tx_\mu \star \phi \star   \tx _\mu - \mu^2 \phi + 2 \lambda   
\, \phi\star\phi^\dag\star\phi = 0 . \nonumber \\  
\eeqa  
Specializing now on the case $\Omega=1$, the equation of motion simplifies to  
\beqa  
\label{eqmvt-complex3}  
\frac 12 (\tx^2 \star \phi + \phi\star \tx^2)   
- \mu^2 \phi + 2 \lambda   
\, \phi\star\phi^\dag\star\phi = 0,  
\eeqa    
For simplicity reasons, we restrict ourselves to $D=2$. 
The adaptation of this analysis to the case $D=4$ is straightforward, as will
be explained in the sequel.

We now look for solutions of \eqref{eqmvt-complex3} under the form
\beqa  
\label{ansatz}  
v(x)= a f_{m_0n_0}(x)
\eeqa  
where $m_0, n_0$ are fixed integers, $a\in\CC^\ast$ and $f _{m_0n_0}(x)$ are
elements of the matrix base (see Appendix \ref{matrix}). 
The equation of motion \eqref{eqmvt-complex3} writes  in the matrix base
\beqa  
\label{eqmvt-matrice}  
\frac{4}{\theta} (m+n+1)\phi_{mn} - \mu^2 \phi_{mn} + 2 \lambda\,  
\phi_{mk} \phi^\dag_{kl} \phi_{ln}=0.\   
\eeqa  
Inserting now in \eqref{eqmvt-matrice} the ansatz \eqref{ansatz}, written as  
\beqa  
\label{ansatz-bis}  
\phi_{mn}=a\, \delta_{mm_0} \delta_{nn_0},\mbox{ with } a\in \CC^\ast  
\eeqa  
one has  
\beqa  
a\Big(\frac{4}{\theta} (m_0+n_0+1)-\mu^2+2 \lambda\, |a|^2 \Big) =0.\   
\eeqa  
This implies  
\beqa  
\label{a}  
|a|^2 = \frac{1}{\lambda \theta} \left( \frac{\mu^2 \theta}{2} - 2(m_0+n_0+1)\right),
\eeqa  
so that consistency requires
the following condition on the mass (or equivalently on the indexes $m_0$ and $n_0$)  
\beqa  
\label{cond}  
\mu^2>\frac{4}{\theta}(m_0+n_0+1).  
\eeqa  
  
Thus  all the functions proposed in \eqref{ansatz} with $a$ satisfying 
\eqref{a} are solutions of the equation of motion if the condition 
\eqref{cond} is verified. We denote  
$$p_C=\lfloor \frac{\mu^2\theta}{4}-1 \rfloor$$ 
 (where $\lfloor . \rfloor$ is the integer part).  
If $p_C$ is negative, the only possibility is the trivial solution. If $p_C$ is 
positive, the solutions indexed by $m_0$ and $n_0$ have to satisfy the 
constraint $m_0+n_0\leq p_C$. The number of solutions of the form \eqref{ansatz} 
is $\sum_{k=0}^{p_C}(p_C-k+1)=\frac{(p_C+1)(p_C+2)}{2}$. We will go further with the 
interpretation of these observations in the next subsection,  where a similar description can be done.  
  
At this point, we make the following observation. The transformation  
\begin{align}  
\phi_{m,n}&\mapsto \phi'_{m,n}=\phi_{m-1,n}\quad(\text{with}\quad\forall n\in\mathbb N\quad \phi'_{0,n}=0)\nonumber\\  
\mu^2&\mapsto \mu^2+\frac{4}{\theta}
\label{sym1}  
\end{align}  
is a symmetry of the equation of motion \eqref{eqmvt-complex3}. 
Let  a solution $v_{mn}=a(\mu^2)\, \delta_{mm_0} \delta_{nn_0}$ and let this symmetry act on it: $v'_{m,n}=a(\mu^2-\frac{4}{\theta})\, \delta_{m,m_0+1} \delta_{n,n_0}$, then $v'_{m,n}$ is also a solution of the equation of motion. As the complex conjugation is also a symmetry of \eqref{eqmvt-complex3}, we find that the composition  
\begin{align}  
\phi_{m,n}&\mapsto \phi'_{m,n}=\phi_{m,n-1}\quad(\text{with}\quad\forall m\in\mathbb N\quad \phi'_{m,0}=0)\nonumber\\  
\mu^2&\mapsto \mu^2+\frac{4}{\theta}\label{sym2}  
\end{align}  
is also a symmetry. We notice that with these both transformations, all the solutions \eqref{ansatz} of equation of motion can be derived from a single one, $\sqrt{\frac{2}{\lambda\theta}\big(\frac{\mu^2\theta}{4}-1\big)}f_{00}(x)$.

 
\medskip 
  
For $\Omega\neq 1$, the equation of motion \eqref{eqmvt-complex2} writes in the matrix base  
\beqa  
\label{eqmvt-om}  
\frac{2}{\theta} (1+ \Omega^2) (m+n+1)\phi_{mn} - \frac{2}{\theta}  (1-  
\Omega^2)\sqrt{(m+1)(n+1)}\phi_{m+1, n+1}\nonumber\\  
- \frac{2}{\theta} (1-  
\Omega^2)\sqrt{mn} \phi_{m-1,n-1} - \mu^2 \phi_{mn} + 2 \lambda\,  
\phi_{mk} \phi^\dag_{kl} \phi_{ln}=0.\   
\eeqa  
All the solutions of the form \eqref{ansatz} do not verify this new equation
\eqref{eqmvt-om} if $\Omega\neq 1$. Nevertheless, it is possible to find some
solutions of \eqref{eqmvt-om}, for instance the sum of two elements of the matrix
base. 
 The full analysis of the case $\Omega\ne 1$ deserves further investigation
 which goes beyond the scope of this paper. 
 
\subsection{The real-valued scalar field theory}  
  
In the case of a real field $\phi$, the equation of motion derived from \eqref{act-real} writes  
\beqa  
\label{eqmvt-real}  
-\partial^2\phi+\Omega^2\widetilde{x}^2\phi-\mu^2\phi+4\lambda\phi\star\phi\star\phi=0,  
\eeqa  
which for $\Omega=1$ can be rewritten as
\beqa  
\label{eqmvt}  
\frac 12 (\tx^2 \star \phi + \phi\star \tx^2)   
- \mu^2 \phi + 4 \lambda   
\, \phi\star\phi\star\phi = 0.  
\eeqa  
We now look for solutions of \eqref{eqmvt} whose form is given by a similar
ansatz to the one of the complex case \eqref{ansatz}. Note however that the
vacuum must now be consistent with the reality condition $v^\dag(x)=v(x)$. We
put 
\beqa  
\label{ansatz-real}  
v(x)= a_{m_0} f_{m_0m_0}(x),\  m_0\in \NN.  
\eeqa  
Note that there is no Einstein convention of summation ($m_0$ is fixed). The equation of motion \eqref{eqmvt} can be re-expressed in the matrix base  
\beqa  
\label{eqmvt-matrix-real}  
\frac{4}{\theta} (m+n+1)\phi_{mn} - \mu^2 \phi_{mn} + 4 \lambda\,  
\phi_{mk} \phi_{kl} \phi_{ln}=0,\   
\eeqa  
and the ansatz  
\beqa  
\label{ansatz-bis-real}  
\phi_{mn}=a_{m_0}\, \delta_{mm_0} \delta_{nm_0},\mbox{ with } a_{m_0}\in \mathbb R^\ast.  
\eeqa  
By insertion of \eqref{ansatz-bis-real} in \eqref{eqmvt-matrix-real}, one finds  
\beqa  
a_{m_0}\Big(\frac{4}{\theta} (2m_0+1)-\mu^2+4 \lambda\, a_{m_0}^2 \Big) =0.\   
\eeqa  
As a consequence, $a_{m_0}$ has to satisfy  
\beqa  
\label{a-real}  
a_{m_0}^2 = \frac{1}{\lambda \theta} \left( \frac{\mu^2 \theta}{4} - 2m_0-1\right),  
\eeqa  
and the mass  
\beqa  
\label{cond-real}  
\mu^2>\frac{4}{\theta}(2m_0+1).  
\eeqa  
Now, upon setting 
\beqa 
\label{p} 
p=\lfloor \frac{\mu^2\theta}{8}-\frac 12 \rfloor 
\eeqa 
(the counterpart of $p_C$ for the real-valued theory), the discussion proceeds
along the same lines as in the previous subsection.
Namely, if $p$ is negative, no index can satisfy the 
 condition \eqref{cond-real}.  
If $p$ is positive, there are $(p+1)$ solutions of the form
\eqref{ansatz-real} satisfying the constraint \eqref{cond-real}.  
Notice that the counterpart of the symmetries \eqref{sym1} and \eqref{sym2} is now  
\begin{align}  
\phi_{m,n}&\mapsto \phi'_{m,n}=\phi_{m-1,n-1}\quad(\text{with}\quad\forall m,n\in\mathbb N\quad \phi'_{0,n}=0\quad\text{and}\quad \phi'_{m,0}=0)\nonumber\\  
\mu^2&\mapsto \mu^2+\frac{8}{\theta}\label{sym3}  
\end{align}  
so that all the vacua of the form \eqref{ansatz-real} are derived from $\sqrt{\frac{2}{\lambda\theta}\big(\frac{\mu^2\theta}{8}-\frac 12\big)}f_{00}(x)$.

\medskip 
 
Let us now look for more general solutions of type: 
\beqa  
\label{ansatz-real-gen}  
v(x)= \sum_{k=0}^\infty a_{k} f_{kk}(x),\  (a_k)\in\mathbb R^{\mathbb N}.  
\eeqa  
The equation of motion \eqref{eqmvt} then leads to the following condition on the coefficients $a_k$  
\beqa  
\label{a-real-gen}  
a_k=0\quad\text{or}\quad a_{k}^2 = \frac{1}{\lambda \theta} \left( \frac{\mu^2 \theta}{4} - 2k-1\right).  
\eeqa  
Owing to the analysis given above, one readily infers that the sum involved in
\eqref{ansatz-real-gen} cannot run to infinity simply because one must have $k\leq p$. 
 
Therefore,   $\sum_{k=0}^p a_{k} f_{kk}(x)$ represents a new set of solutions
for the equation of motion which cannot be derived from a single one under the symmetry \eqref{sym3}.
One now has for the coefficients of $v\star v\star\phi$ and $v\star\phi\star v$:  
\begin{align}  
(v\star v\star\phi)_{mn}&=\big(\sum_{k=0}^p a_k^2\delta_{mk}\big)\phi_{mn},\nonumber\\  
(v\star\phi\star v)_{mn}&=\big(\sum_{k,l=0}^p a_ka_l\delta_{mk}\delta_{nl}\big)\phi_{mn}.  
\end{align}  
Let us now check if these solutions are minima for the action. 
The quadratic part of the action \eqref{act-real} writes 
\begin{align}  
S_{quadr}=2\pi\theta\Big(\frac{2}{\theta}(m+n+1)-\frac{\mu^2}{2}+2\lambda\sum_{k=0}^p a_k^2(\delta_{mk}+\delta_{nk})+ 2\lambda\sum_{k,l=0}^p a_ka_l\delta_{mk}\delta_{nl}\Big)\phi_{nm}\phi_{mn}.\label{act-reexpr}  
\end{align}  
The propagator $C_{mn,kl}$ is diagonal  
$$C_{mn,kl}=C_{mn}\delta_{m l}\delta_{nk}.$$  
In order to have a minimum, one thus needs 
$C_{mn}$ to be positive, for all $m,n\in\mathbb N$. From \eqref{act-reexpr} 
one has  
\begin{align}  
C_{mn}^{-1}=\alpha_{mn}+4\lambda\pi\theta\sum_{k=0}^p a_k^2(\delta_{mk}+\delta_{nk})+ 4\lambda\pi\theta\sum_{k,l=0}^p a_ka_l\delta_{mk} \delta_{nl},  
\end{align}  
where  
\begin{align}  
\alpha_{mn}=4\pi(m+n+1)-\mu^2\pi\theta,\mbox{ and } a_k^2=0\quad\text{or}\quad a_k^2=-\frac{\alpha_{kk}}{4\lambda\pi\theta}.  
\end{align}  
  
For $p\ge 0$, one has to distinguish between the following cases.  
\begin{itemize}  
\item $m>p$ and $n>p$:  
\begin{align}  
C^{-1}_{mn}=\alpha_{mn}=4\pi(m+n-2(\frac{\mu^2\theta}{8}-\frac 12))>0.  
\end{align}  
\item $m\leq p$ and $n>p$:  
if $a_m^2=0$ then  
\begin{align}  
C^{-1}_{mn}=\alpha_{mn}=4\pi(m+n-2(\frac{\mu^2\theta}{8}-\frac 12)).\label{condpropag}  
\end{align}  
In order not to have for certain value of $n$ ($p< n\leq 2p-m$) $C_{mn}^{-1}<0$, one needs that $a_m^2=-\frac{\alpha_{mm}}{4\lambda\pi\theta}$. In this case, we have  
\begin{align}  
C^{-1}_{mn}=\alpha_{mn}-\alpha_{mm}=4\pi(n-m)>0.  
\end{align}  
\item $m>p$ and $n\leq p$: is treated along the same lines as above. $C^{-1}_{mn}=4\pi(m-n)>0$ because we assumed that $\forall k\in \{0,..,p\}$, $a_k^2=-\frac{\alpha_{kk}}{4\lambda\pi\theta}$.  
\item $m\leq p$ and $n\leq p$:  
\begin{align}  
C^{-1}_{mn}=\alpha_{mn}-\alpha_{mm}-\alpha_{nn}+\sqrt{\alpha_{mm}\alpha_{nn}}=-\alpha_{mn}+\sqrt{\alpha_{mm}\alpha_{nn}}\geq 0,  
\end{align}  
and $C^{-1}_{mn}=0$ holds if and only if $m=n=\frac{\mu^2\theta}{8}-\frac 12\in\mathbb N$.  
\end{itemize}  
  
From the above analyze, one can conclude that one has a positive defined propagator just for a single solution. This solution, which is a minimum of the action \eqref{act-real}, corresponds to  
\begin{align}  
v(x)=\sum_{k=0}^p a_k f_{kk}(x)\label{zevacuum}  
\end{align}  
where $a_{k}^2 = \frac{1}{\lambda \theta} \left( \frac{\mu^2 \theta}{4} - 
  2k-1\right)$.

For $\phi(x)=\sum_{k,l=0}^\infty\phi_{mn}f_{mn}(x)$ a solution of equation of motion \eqref{eqmvt}, we compute the value of the action \eqref{act-real} in the matrix base
\begin{align}
S[\phi]=2\pi\sum_{k,l=0}^\infty\left( m+n-2\big(\frac{\mu^2\theta}{8}-\frac 12\big)\right)|\phi_{mn}|^2.
\end{align}
So for $p<0$,
\begin{align}
S[\phi]\geq S[0]=0,\label{actmin}
\end{align}
and for $p\geq 0$ and $v(x)$ the vacuum \eqref{zevacuum},
\begin{align}
S[v]=-\frac{8\pi}{\lambda \theta} \sum_{k=0}^p (\frac{\mu^2\theta}{8}-\frac12-k)^2<0.
\end{align}

At this point, a remark is to be done. In commutative QFT or in 
non-commutative QFT without the harmonic term, one has a phenomena of 
spontaneously symmetry breaking as soon as the mass parameter is taken to be 
negative. For the models considered here this is not the case anymore. Indeed, 
if the mass parameter does not go beyond a certain limit $\mu^2=\frac{4}{\theta}$, then from \eqref{actmin}, we see that $\phi(x)=0$ is the global minimum of  the action \eqref{act-real}, as $p$ is negative. So the harmonic 
 term will prevent the phenomena of spontaneously symmetry 
breaking from happening. 

When the mass parameter exceeds this critical value (i.e. $p\geq 0$), one can see from \eqref{condpropag} that $\phi(x)=0$ is no longer a local minimum of the action. Therefore one has to consider a non-trivial vacuum $v(x)$, as for example \eqref{zevacuum}. Note that \eqref{zevacuum} corresponds to a different solution when changing the value of the limit 
parameter $p$. Let us now stress on some of the features of these vacuum configurations. Owing again to the properties of the matrix base (see again  Appendix
 \ref{matrix}), it can be realized that \eqref{zevacuum} does not vanish for
 $x=0$ while it decays at infinity (as a {\it finite} linear combination of the Schwartz functions $f_{mn}(x)$).

Let us further indicate that these results extend in $D=4$ also. Indeed, one
just has to replace the indices $m$ by $(m_1, m_2)\in \NN^2$ and  the number
$m$ by $m_1+m_2$. In four dimensions, it is well known that when computing
radiative corrections, the mass parameter of a scalar field becomes huge
(because of the quadratic divergence). In order to get a low value for the
renormalized mass, one may thus consider a non-commutative scalar field
theory with harmonic term, a negative mass term and a non-trivial
vacuum. It is further possible to choose \eqref{zevacuum} as this
non-trivial vacuum, since the special value $\Omega=1$ is stable under the renormalization group.

\medskip 
 
Moreover we also exhibit in the next subsection the following class of
solutions of the equation of motion. If one has a configuration 
$v(x)$ which satisfies   
  \beqa  
  v\star v=-\frac{1}{4\lambda}\widetilde{x}^2+\frac{\mu^2}{4\lambda}  
  \eeqa  
  (see \eqref{contrainte}, \eqref{om} and \eqref{rez}) then $v(x)$ will also
  be a solution
  of the equation of motion. 
As already stated in subsection  
  \ref{goldstone} this type of equation can be easily shown to have non-trivial  
  solutions using again the matrix base.   

\medskip

We have thus pointed out in this subsection the existence of a non-trivial 
vacuum $v(x)$. Such a vacuum can be used in the next section, where we 
consider fluctuations of fields around such $v(x)$. 

\subsection{Spontaneous symmetry breaking for the linear sigma model}  
\label{goldstone}

  
 
 
As an warming up issue, consider again the action \eqref{act-real}  for a real $\phi$ field. This   
action has a discrete symmetry:  
\beqa  
\label{discret}  
\phi\to -\phi.  
\eeqa  
As usual, assume that  the system is near one of its minima $v(x)$. 
Upon setting 
\beqa  
\phi(x)= v(x)+\sigma (x)
\eeqa  
in \eqref{act-real}, one obtains 
the Lagrangian in terms of the $\sigma$ field 
\begin{align}  
S=\int d^4x  \big( 
& (\frac 12\partial_\mu\sigma)\star(\partial_\mu\sigma)+  
\frac{\Omega^2}{2}\widetilde{x}^2\sigma\sigma -\frac{\mu^2}{2}\sigma \sigma+4\lambda v \star v\star\sigma \star\sigma  \nonumber\\  
&+2 \lambda v\star \sigma \star\ v\star\sigma + 4\lambda v \star \sigma\star  
\sigma\star \sigma + \lambda \sigma \star \sigma \star\sigma\star\sigma \big).  
\end{align}  
(where we have used \eqref{simplu}). Note that, as in the commutative case, the symmetry \eqref{discret} has  
disappeared. The situation is exactly the same in the case of a complex field \eqref{act-complex}.

\medskip 
  
We now consider the linear sigma model build 
from the renormalizable  
scalar action \eqref{act-real}, assuming again $\Omega=1$. 
Additional considerations for $\Omega\ne 1$ 
as well as for the case of a complex valued field 
will be given at the end of this section.  
  
The action involves $N$ valued fields $\phi_i$ and is given by
\beqa  
\label{act-sigma}  
S_\sigma=\int d^4x\Big( \frac 12(\partial_\mu\phi_i)\star(\partial_\mu\phi_i)  
+ \frac 12\widetilde{x}^2\phi_i \phi_i-\frac{\mu^2}{2}\phi_i \phi_i  
+\lambda\phi_i \star\phi_i\star\phi_j \star\phi_j\Big)  
\eeqa  
The action  
above is invariant under the action of the orthogonal group $O(N)$  
(as it is also the case  in the absence of the harmonic  term, see  
\cite{gold1, gold11, gold2}).  
  
Let 
\beqa  
\label{vev}  
<\Phi>=(0, \ldots, 0, v(x))  
\eeqa  
a non-zero vacuum expectation value, 
where $v(x)$ is some minimum obtained from the equation of motion, as  analyzed in the  
previous section.   
  
Then shifting $\Phi$ to $<\Phi>+\delta\Phi$ with
\beqa  
\label{shift}  
\delta \Phi=(\pi_1,\ldots, \pi_{N-1},\sigma (x))  
\eeqa  
one obtains from \eqref{act-sigma}  
\begin{align}  
\label{act-dev}  
S_\sigma=\int d^4x &\Big( \frac 12(\partial_\mu\pi_i) \star(\partial_\mu\pi_i)+ \frac 12\widetilde{x}^2\pi_i \pi_i-\frac{\mu^2}{2}\pi_i \pi_i+2\lambda v \star v\star\pi_i \star\pi_i+\lambda\pi_i \star\pi_i\star\pi_j \star\pi_j \nonumber\\  
& +\frac 12(\partial_\mu\sigma) \star(\partial_\mu\sigma)+ \frac 12\widetilde{x}^2\sigma \sigma-\frac{\mu^2}{2}\sigma \sigma+4\lambda v \star v\star\sigma \star\sigma +2\lambda v\star \sigma \star\ v\star\sigma \nonumber\\  
&  +2\lambda \sigma \star v\star\pi_i \star\pi_i+2\lambda v \star \sigma\star\pi_i \star\pi_i +2\lambda \sigma \star \sigma\star\pi_i \star\pi_i\nonumber\\  
&+4\lambda v\star\sigma\star\sigma\star\sigma+\lambda\sigma\star\sigma\star\sigma\star\sigma\Big).  
\end{align}  
Consider closer the part of the action \eqref{act-dev} 
quadratic in the fields  
$\pi$:   
\beqa  
\label{muie}
\int d^4 x \big(\frac 12\widetilde{x}^2\pi_i\pi_i-\frac{\mu^2}{2}\pi_i\pi_i+2\lambda  
v\star v\star\pi_i\star\pi_i\big).  
\eeqa  
In the absence of the harmonic term, the linear sigma model supports a
constant non-zero vacuum configuration leading to the appearance of $N$
massless fields $\pi$. Indeed, the second and the third term in \eqref{muie}
balance each other. This is an obvious analogue of the Goldstone theorem at the
classical level which has been further verified to the one and resp. two loop order
in \cite{gold1, gold11}  
and resp. \cite{gold2}.  

When the harmonic term is included in \eqref{muie} the situation changes
substantially. Indeed, in view of the discussion for the scalar field theory
presented above, constant non-zero vacuum configurations are no longer
supported by the action. Thus, the cancellation of the $\pi$ mass term does
not occur automatically and must be reconsidered carefully. The attitude we
adopt here is to mimic one of the main feature of the Goldstone theorem of
commutative field theory.

   
 

We  thus investigate whether or not one can find a vacuum $v(x)$ which makes that  
\beqa   
\label{imp}  
&&\int d^4x \, \big(\frac 12\widetilde{x}^2\pi_i\pi_i-\frac{\mu^2}{2}\pi_i\pi_i+2\lambda  
 v\star v\star\pi_i\star\pi_i\big)=  \nonumber\\  
&&\int d^4x \, \left( \frac 12\widetilde{x}^2\pi_i\pi_i-\frac{\mu^2}{2}(\pi_i\star\pi_i)+2\lambda  
 (v\star v)(\pi_i\star\pi_i)\right)=  
\int d^4x \left(  
 \frac{\Omega'^2}{2}\widetilde{x}^2\pi_i\pi_i + ... \right)\nonumber\\  
\eeqa  
Note that by the dots on the RHS of the second line of \eqref{imp} we mean some eventual kinetic terms. If such  
a statement holds this means that one just has some  
harmonic type of term for the fields $\pi$; moreover, these fields  
would be   
non-massive. It is this what we propose as a corresponding Goldstone theorem  
in our case.  
Furthermore  we also allow the possibility $\Omega'=0$ (the fields $\pi$ are non-massive and they have no harmonic term  
neither).   
  
Moreover, note that this type of masslessness constraint for the $\pi$ fields  
is the most general one can impose here (also introducing a harmonic   
like free parameter $\Omega'$).   
  
Finally, looking at the LHS of the second line of \eqref{imp} one sees that  
all the terms contain a $\pi\star\pi$ product except for the first one. In  
order to be able to factorize this  $\pi\star\pi$ product, we now re-express  
the first term of the LHS also.  
  
Using \eqref{simplu} and \eqref{ajutor2} one has   
\begin{align}  
\label{final}  
\int  d^4x \, \widetilde{x}^2\pi_i\pi_i=   
\int  d^4x \, (\widetilde{x}^2\pi_i)\star\pi_i=  
\int  d^4x  
\left(\widetilde{x}^2(\pi_i\star\pi_i)-\partial^\mu \pi_i\partial_\mu \pi_i-2i\pi_i\widetilde{x}_\mu\partial_\mu\pi_i\right).  
\end{align}  
Using now \eqref{int} one cancels out the last term in \eqref{final}. All this  
becomes   
\begin{align}  
\label{final2}  
\int  d^4x \, \widetilde{x}^2\pi_i\pi_i= \int  d^4x  
\left(\widetilde{x}^2(\pi_i\star\pi_i)-\partial^\mu \pi_i\partial_\mu \pi_i\right).  
\end{align}  
Note that this way of writing may be misleading in the sense that the kinetic  
term of the $\pi$ fields seem to cancel out with the second term in the RHS of  
\eqref{final2}. However this is just because of the particular way  
\eqref{final2} of writing down the harmonic term at $\Omega=1$.

One now introduces \eqref{final2} in \eqref{imp}. Writing in the same away as  
above the RHS of \eqref{imp} and leaving aside the kinetic terms, one is  
finally able to factorize the product $\pi\star \pi$ to get  
the following constraint for the vacuum $v$  
\begin{align}  
v\star v=-\frac{\omega^2}{4\lambda}\widetilde{x}^2+\frac{\mu^2}{4\lambda}  
\label{contrainte}  
\end{align}  
where  
\beqa  
\label{om}  
\omega^2=1-\Omega'^2.  
\eeqa  
We have thus proven that this constraint is equivalent to the constraint  
\eqref{imp} that we imposed on the kinetic terms of the $\pi$  
fields. Moreover, note that \eqref{contrainte} presents non-trivial solutions  
$v(x)$ as can be seen for example in the matrix base.  
  
\medskip  
  
We now prove that a non-trivial vacuum $v(x)$ satisfying \eqref{contrainte} is a solution  
of the equation of movement \eqref{eqmvt} if and only if one has  
$\Omega'=0$. Indeed,  inserting \eqref{contrainte} in \eqref{eqmvt} one has  
\beqa  
\frac 12(\tx^2 \star v + v\star \tx^2) - \mu^2 \star v + 2\lambda \, v \star  
(-\frac{\omega^2}{4\lambda}\widetilde{x}^2+\frac{\mu^2}{4\lambda}) + 2\lambda  
(-\frac{\omega^2}{4\lambda}\widetilde{x}^2+\frac{\mu^2}{4\lambda}) \star v =0.  
\eeqa  
which rewrites as  
\beqa  
\label{ec}  
\frac 12 (1 - \omega^2) ( v\star \tx^2 + \tx^2 \star v)=0.  
\eeqa  
If one now requires a non-trivial vacuum $v(x)$, then using \eqref{diag}, one has that the coefficients in the matrix base of $v\star  
\tx^2 + \tx^2 \star v$ are $\frac{8}{\theta}(m+n+1)v_{mn}$ and as a consequence, $v\star  
\tx^2 + \tx^2 \star v \ne 0$.  This implies that the only solution of  
\eqref{ec} is  
\beqa  
\label{rez}  
\omega=1  
\eeqa  
which, by \eqref{om} leads to  
\beqa  
\label{dezastru}  
\Omega'=0,  
\eeqa  
as already stated above.  
  
\medskip  
  
However, \eqref{dezastru} considerably simplifies \eqref{imp} which, once  
introduced in the action \eqref{act-dev} leaves, for the $\pi$ fields part,  
only the $\pi^4$ interaction term. Obviously, this is not physically  
satisfying.  
  
 
Finally, let us remark that this situation is imposed by the cancellation  
\eqref{dezastru}, which on its turn is a consequence of the constraint  
\eqref{contrainte} for a vacuum $v(x)$ (solution of the equation of
motion). So it seems that the condition \eqref{contrainte}, coming from the Goldstone
theorem in commutative theories, cannot be imposed for this type of models.
  
As stated in the beginning of this section, all these calculations are made in  
the case of a set of real fields $\phi$ and for the particular value  
$\Omega=1$. If one allows other values of the parameter $\Omega$ and also  
considers complex fields $\phi$, the situation is more intricate. A possible  
way of approach is to combine the two constraints ({\it i. e.} the equation of motion and  
the masslessness \eqref{imp} of the fields $\pi$) into some stronger constraint  
for the vacuum $v(x)$, constraint which finally has to be checked for  
solutions.

\section{Concluding remarks}  
\resetequ  
\label{discutii}  
  
We have thus analyzed in this paper the spontaneous symmetry breaking of the
non-commutative scalar model with harmonic term, and we found that the mass
value $\mu^2=\frac{4}{\theta}$ has a particular importance. For the real case,
for $\mu^2<\frac{4}{\theta}$, the value $\phi(x)=0$ is the global minimum of
the action. For $\mu^2\ge \frac{4}{\theta}$, the value  $\phi(x)=0$  is no
longer a local minimum, so the theory acquires a non-trivial vacuum; a local
minimum \eqref{zevacuum} of the action was found. A further line of work is to
develop the theory around this vacuum and to study its renormalizability and
its renormalization group flows, as it seems to provide low values for the renormalized mass.

We have also analyzed the spontaneous  
symmetry breaking for a corresponding linear sigma model  
with $N$ scalar fields and a harmonic term present in the  
action for each of these $N$ fields. Even though this seems the most natural  
way to construct such a linear sigma model, one cannot {\it a priori} state a  
conclusion with respect to the renormalizability of this model. Moreover, since one has to deal with vacuums $v$ which have a non-trivial  
dependence on the space-time position $x$, one can argue on the interpretation  
of phenomena like spontaneous symmetry breaking or the Goldstone  
theorem. What we have achieved in this paper is a calculation of these usual notions  
of commutative classical field theory in the framework of Grosse-Wulkenhaar like models.

Finally, let us end this paper by reminding the existence of a second class of  
non-commutative models, called ``covariant models''. Here  
one can include the non-commutative Gross-Neveu model or the  
Langmann-Szabo-Zarembo model \cite{LSZ}. These models also were proven to be  
renormalizable \cite{V} and their one-loop $\beta$-function was computed \cite{jc1}. Moreover, their parametric representation was  
implemented in \cite{param2} (see \cite{param} for a general review) and the Mellin representation of their Feynman  
amplitudes (as well as for the Grosse-Wulkenhaar model) was obtained in  
\cite{mellin}.  
\footnote{For some general reviews on latest developments on  
  renormalizability of non-commutative quantum field theories, one may report  
  to \cite{raimar, vince}.}

\appendix  
  
\section{The matrix base}  
\label{matrix}  
\resetequ

In this appendix we present the definition and some useful properties of the matrix  
base of the Moyal space. For more details, see for example   
\cite{matrix1} or \cite{matrix2}.  
  
Let $f_{mn}$, $m,n\in \NN^\frac D2$ the set of Schwartz functions forming the
matrix base, to be given below. Let  
\beqa  
\label{Hl}  
H=\sum_{\ell=1}^\frac D2 H_\ell,\ \ H_\ell=\frac 12 (x_{2\ell -1}^2 + x_{2\ell}^2), \mbox{ for } \ell=1,\ldots,  
\frac D2  
\eeqa  
Furthermore, let   
\beqa  
\label{f00}  
f_{00}(x)=2^\frac D2 e^{-\frac{2}{\theta}H}.  
\eeqa  
This verifies 
\beqa  
\label{prop00}  
f_{00}\star f_{00} = f_{00}.  
\eeqa  
One also defines the operators  
\beqa  
\label{creatie}  
a_\ell =\frac{1}{\sqrt{2}} (x_{2\ell-1} + i x_{2\ell}),\  \
\bar a_\ell =\frac{1}{\sqrt{2}} (x_{2\ell-1} - i x_{2\ell}) .  
\eeqa  
together with  
\beqa  
\label{def}  
f_{mn}(x)=\frac{1}{\sqrt{m!n!\theta^{m+n}}}\, \bar a^{\star m} \star f_{00} \star a^{\star n}.  
\eeqa  
These functions diagonalize the Hamiltonian \eqref{Hl}:  
\beqa  
\label{diag}  
H\star f_{mn} = \theta (m + \frac{1}{2})f_{mn} ,\ \  
f_{mn}\star H=\theta(n+\frac{1}{2})f_{mn}.  
\eeqa  
Some useful properties are
\beqa  
\label{proprietati}  
f_{mn}^\dag&=&f_{nm},\nonumber\\  
f_{mn}\star  f_{kl}(x)&=&\delta_{nk}f_{ml}(x).  
\eeqa  
  
\medskip  
  
Finally, let us give, in $D=2$, the functions  
\beqa  
 f_{10}(x)&=&2\sqrt{\frac{2}{\theta}}(x_1-ix_2)e^{-\frac{x^2}{\theta}}\nonumber\\   
f_{01}(x)&=&2\sqrt{\frac{2}{\theta}}(x_1+ix_2)e^{-\frac{x^2}{\theta}}  
\eeqa  
  
\medskip  
  
{\textbf{Acknowledgments}}: We wish to thank   
R. Gurau, J. Magnen, T. Masson, V. Rivasseau, F. Vignes-Tourneret and  R. Wulkenhaar   
for  
useful discussions during the preparation of this work.


\begin{thebibliography}{99}  
  
\bibitem{DN}  
M.~Douglas, N.~Nekrasov,  
``Noncommutative field theory,''  
Rev. Modern Physics {\bf 73}, 977-1029 (2001). 
  
\bibitem{Szabo}  
R.~J.~Szabo,  
  ``Quantum field theory on noncommutative spaces,''  
  Phys.\ Rept.\  {\bf 378}, 207 (2003)  
  [arXiv:hep-th/0109162].  
  
\bibitem{book-connes}  
A. Connes, ``G\'eometrie non commutative'', InterEditions, Paris (1990).  
  
\bibitem{string1}   
A.~Connes, M. R.~Douglas, A.~Schwarz, 
``Noncommutative Geometry and Matrix Theory: Compactification on Tori,''  
JHEP {\bf 9802}, 3-43 (1998). 
  
\bibitem{string2}  
N.~Seiberg, E.~Witten, 
``String theory and noncommutative geometry,''  
JHEP {\bf 9909}, 32-131 (1999)  
[arXiv:hep-th/9908142].  
  
\bibitem{hall1}   
L.~Susskind, ``The Quantum Hall Fluid and Non-Commutative Chern Simons Theory,''  
[arXiv:hep-th/0101029]. 
  
\bibitem{hall2}   
A. P.~Polychronakos, 
``Quantum Hall states on the cylinder as unitary matrix Chern-Simons theory,''  
JHEP, {\bf 06}, 70-95 (2001). 
  
  
\bibitem{hall3}   
S.~Hellerman, M.~Van Raamsdonk, 
``Quantum Hall physics equals noncommutative field theory,'' 
JHEP {\bf 10}, 39-51 (2001). 
  
  
\bibitem{GW1}  
H.~Grosse and R.~Wulkenhaar,  
``Power-counting theorem for non-local matrix models and renormalization,''  
 Commun.\ Math.\ Phys. {\bf 254}, 91-127 (2005). 
  
\bibitem{GW2}   
H.~Grosse and R.~Wulkenhaar,  
``Renormalizationof $\phi^4$-theory on noncommutative ${\mathbb R}^4$ in the matrix base,''  
 Commun.\ Math.\ Phys. {\bf 256}, 305-374 (2005). 
  
\bibitem{GW5}  
 H.~Grosse and R.~Wulkenhaar,  
 ``Renormalisation of phi**4-theory on non-commutative R**4 to all orders,''  
  Lett.\ Math.\ Phys.\  {\bf 71}, 13 (2005).  
  
\bibitem{GW3}  
V.~Rivasseau, F.~Vignes-Tourneret and R.~Wulkenhaar,  
``Renormalization of noncommutative $\phi^{\star4}_4$-theory by multi-scale analysis,''  
Commun. Math. Phys. {\bf 262}, 565-594 (2006)  
  
\bibitem{GW4}  
R.~Gurau, J.~Magnen, V.~Rivasseau and F.~Vignes-Tourneret,  
  ``Renormalization of non-commutative phi**4(4) field theory in x space,''  
  Commun.\ Math.\ Phys.\  {\bf 267}, 515 (2006)  
  [arXiv:hep-th/0512271].  
  
\bibitem{param1}  
R.~Gurau and V.~Rivasseau,  
  ``Parametric representation of noncommutative field theory,''  
  Commun.\ Math.\ Phys.\  {\bf 272}, 811 (2007)  
  [arXiv:math-ph/0606030].  
  
  
\bibitem{dimreg}  
R.~Gurau and A.~Tanasa,  
  ``Dimensional regularization and renormalization of non-commutative QFT,''  
  submitted to Annales Henri Poincare,   
  arXiv:0706.1147 [math-ph].  
  
\bibitem{hopf}  
 A.~Tanasa and F.~Vignes-Tourneret,  
  ``Hopf algebra of non-commutative field theory,''  
to be published in J. Noncomm. Geom.,     
arXiv:0707.4143 [math-ph].  
  
  
\bibitem{ultimul}  
  H.~Grosse and R.~Wulkenhaar,  
  ``8D-spectral triple on 4D-Moyal space and the vacuum of noncommutative gauge  
  theory,''  
  arXiv:0709.0095 [hep-th].  
  
  
\bibitem{spectral}  
  A.~H.~Chamseddine, A.~Connes and M.~Marcolli,  
  ``Gravity and the standard model with neutrino mixing,''  
  arXiv:hep-th/0610241.  
  
  
  
\bibitem{beta1}  
H.~Grosse and R.~Wulkenhaar,  
  ``The beta-function in duality-covariant noncommutative phi**4 theory,''  
  Eur.\ Phys.\ J.\  C {\bf 35}, 277 (2004)  
  [arXiv:hep-th/0402093].  
  
\bibitem{beta23}  
M.~Disertori and V.~Rivasseau,  
  ``Two and three loops beta function of non commutative phi(4)**4 theory,''  
  Eur.\ Phys.\ J.\  C {\bf 50}, 661 (2007)  
  [arXiv:hep-th/0610224].  
  
\bibitem{beta}  
 M.~Disertori, R.~Gurau, J.~Magnen and V.~Rivasseau,  
  ``Vanishing of beta function of non commutative phi(4)**4 theory to all  
  orders,''  
  Phys.\ Lett.\  B {\bf 649}, 95 (2007)  
  [arXiv:hep-th/0612251].  
  
\bibitem{carte}   
V.~Rivasseau, ``From perturbative to Constructive Field Theory,''  
Princeton University Press, (1991)  
  
\bibitem{constructiva1}  
  V.~Rivasseau,  
  ``Constructive Matrix Theory,''  
  arXiv:0706.1224 [hep-th].  
  
\bibitem{constructiva2}  
  J.~Magnen and V.~Rivasseau,  
  ``Constructive $\phi^4$ field theory without tears,''  
  arXiv:0706.2457 [math-ph].  
  
    
\bibitem{gold1}  
  B.~A.~Campbell and K.~Kaminsky,  
  ``Noncommutative linear sigma models,''  
  Nucl.\ Phys.\  B {\bf 606}, 613 (2001)  
  [arXiv:hep-th/0102022].  
  B.~A.~Campbell and K.~Kaminsky,  
  ``Noncommutative field theory and spontaneous symmetry breaking,''  
  Nucl.\ Phys.\  B {\bf 581}, 240 (2000)  
  [arXiv:hep-th/0003137].  
  
  
  
\bibitem{gold11}  
  F.~Ruiz Ruiz,  
  ``UV/IR mixing and the Goldstone theorem in noncommutative field theory,''  
  Nucl.\ Phys.\  B {\bf 637}, 143 (2002)  
  [arXiv:hep-th/0202011].  
  
\bibitem{gold2}  
  Y.~Liao,  
  ``Validity of Goldstone theorem at two loops in noncommutative U(N)  linear  
  sigma model,''  
  Nucl.\ Phys.\  B {\bf 635}, 505 (2002)  
  [arXiv:hep-th/0204032].  
  
\bibitem{higgs}  
F.~J.~Petriello,  
  ``The Higgs mechanism in non-commutative gauge theories,''  
  Nucl.\ Phys.\  B {\bf 601}, 169 (2001)  
  [arXiv:hep-th/0101109].  
  
  
\bibitem{stripe}  
S.~S.~Gubser and S.~L.~Sondhi,  
  ``Phase structure of non-commutative scalar field theories,''  
  Nucl.\ Phys.\  B {\bf 605}, 395 (2001)  
  [arXiv:hep-th/0006119].  
  
\bibitem{altii1}  
  H.~O.~Girotti, M.~Gomes, A.~Y.~Petrov, V.~O.~Rivelles and A.~J.~da Silva,  
  Phys.\ Rev.\  D {\bf 67}, 125003 (2003)  
  [arXiv:hep-th/0207220].  
  
\bibitem{altii2}  
  P.~Castorina and D.~Zappala,  
  ``Nonuniform symmetry breaking in noncommutative lambda phi**4 theory,''  
  Phys.\ Rev.\  D {\bf 68}, 065008 (2003)  
  [arXiv:hep-th/0303030].  
  
\bibitem{matrix1}  
J.~M.~Gracia-Bondia and J.~C.~Varilly,  
  ``Algebras of distributions suitable for phase space quantum mechanics. 1,''  
  J.\ Math.\ Phys.\  {\bf 29}, 869 (1988).  

\bibitem{moyal2}
  J.~C.~Varilly and J.~M.~Gracia-Bondia,
  ``Algebras of distributions suitable for phase-space quantum mechanics. {II}.
  Topologies on the Moyal algebra,''
  J.\ Math.\ Phys.\  {\bf 29}, 880 (1988);
 A. Grossmann, G. Laupias, E. M. Stein, ``An algebra of pseudodifferential
 operators and quantum mechanics in phase space'', Ann. Inst. Fourier {\bf
   18}, 343 (1968).

\bibitem{raimar}  
  R.~Wulkenhaar,  
  ``Field Theories On Deformed Spaces,''  
  J.\ Geom.\ Phys.\  {\bf 56}, 108 (2006).  
  


\bibitem{gauge4}  
  J.~C.~Wallet,  
  ``Noncommutative Induced Gauge Theories on Moyal Spaces,''  
  arXiv:0708.2471 [hep-th], to be published in ``J. Physics: Conf. Series''.   

\bibitem{gauge1}  
A.~de Goursac, J.~C.~Wallet and R.~Wulkenhaar,  
  ``Noncommutative Induced Gauge Theory,''  
Eur. Phys. J. C {\bf 51} (2007) 977  
 [arXiv:hep-th/0703075]. 
  
\bibitem{gauge2}  
H.~Grosse and M.~Wohlgenannt,  
  ``Induced Gauge Theory on a Noncommutative Space,''  
  [arXiv:hep-th/0703169].  
  
  

  
  
\bibitem{gauge5}  
A.~de Goursac, ``On the Effective Action of Noncommutative Yang-Mills Theory,'' 
to be published in J. Physics: Conf. Series. 
  

  
\bibitem{matrix2}  
H.~Grosse and R.~Wulkenhaar,  
  ``Renormalisation of phi**4 theory on noncommutative R**2 in the matrix  
  base,''  
  JHEP {\bf 0312}, 019 (2003)  
  [arXiv:hep-th/0307017].  
  


\bibitem{ls}  
E.~Langmann and R.~J.~Szabo,  
  ``Duality in scalar field theory on noncommutative phase spaces,''  
  Phys.\ Lett.\  B {\bf 533}, 168 (2002)  
  [arXiv:hep-th/0202039].  
  
  
  
  
\bibitem{LSZ}  
E.~Langmann, R. J.~Szabo, K.~Zarembo,  
``Exact solution of quantum field theory on noncommutative phase spaces,''  
JHEP {\bf 0401}, 17-87 (2004).  
  
  
\bibitem{V}   
F.~Vignes-Tourneret, 
``Renormalization of the orientable   
non-commutative Gross-Neveu model,''  
Annales Henri Poincare {\bf 8}, 427 (2007)
  [arXiv:math-ph/0606069].
  
\bibitem{jc1} 
  A.~Lakhoua, F.~Vignes-Tourneret and J.~C.~Wallet, 
  ``One-loop beta functions for the orientable non-commutative Gross-Neveu 
  model,''  to be published in Eur. Phys. J. C 
  [arXiv:hep-th/0701170]. 
 
 
\bibitem{param2}  
  V.~Rivasseau and A.~Tanasa,  
  ``Parametric representation of 'covariant' noncommutative QFT models,''  
  submitted to Commun. Math. Phys. [arXiv:math-ph/0701034].  
  
  
\bibitem{param}  
A.~Tanasa, ``Overview of the parametric representation of renormalizable  
non-commutative field theory'', to be published in J. Physics: Conf. Series,arXiv:0709.2270 [math-ph]. 
  
\bibitem{mellin}  
  R.~Gurau, A.~P.~C.~Malbouisson, V.~Rivasseau and A.~Tanasa,  
  ``Non-Commutative Complete Mellin Representation for Feynman Amplitudes,''
  Lett.\ Math.\ Phys.\  {\bf 81}, 161 (2007),  
  arXiv:0705.3437 [math-ph].  
  
\bibitem{vince}  
  V.~Rivasseau,  
  ``Non-commutative renormalization,''  
  arXiv:0705.0705 [hep-th].  
 

  
\end{thebibliography}
\end{document}